\newif\ifpagetitre            \pagetitretrue
\newtoks\hautpagetitre        \hautpagetitre={\hfil}
\newtoks\baspagetitre         \baspagetitre={\hfil}
\newtoks\auteurcourant        \auteurcourant={\hfil}
\newtoks\titrecourant         \titrecourant={\hfil}

\newtoks\hautpagegauche       \newtoks\hautpagedroite
\hautpagegauche={\hfil\the\auteurcourant\hfil}
\hautpagedroite={\hfil\the\titrecourant\hfil}

\newtoks\baspagegauche \baspagegauche={\hfil\tenrm\folio\hfil}
\newtoks\baspagedroite \baspagedroite={\hfil\tenrm\folio\hfil}

\headline={\ifpagetitre\the\hautpagetitre
\else\ifodd\pageno\the\hautpagedroite
\else\the\hautpagegauche\fi\fi}

\footline={\ifpagetitre\the\baspagetitre
\global\pagetitrefalse
\else\ifodd\pageno\the\baspagedroite
\else\the\baspagegauche\fi\fi}

\vsize=9.0in\voffset=1cm
\looseness=2


\message{fonts,}

\font\tenrm=cmr10
\font\ninerm=cmr9
\font\eightrm=cmr8
\font\teni=cmmi10
\font\ninei=cmmi9
\font\eighti=cmmi8
\font\ninesy=cmsy9
\font\tensy=cmsy10
\font\eightsy=cmsy8
\font\tenbf=cmbx10
\font\ninebf=cmbx9
\font\tentt=cmtt10
\font\ninett=cmtt9

\font\ninesl=cmsl9
\font\eightsl=cmsl8

\font\nineit=cmti9
\font\eightit=cmti8

\skewchar\ninei='177 \skewchar\eighti='177
\skewchar\ninesy='60 \skewchar\eightsy='60

\def\eightpoint{\def\rm{\fam0\eightrm} 
\normalbaselineskip=9pt
\normallineskiplimit=-1pt
\normallineskip=0pt

\textfont0=\eightrm \scriptfont0=\sevenrm \scriptscriptfont0=\fiverm
\textfont1=\ninei \scriptfont1=\seveni \scriptscriptfont1=\fivei
\textfont2=\ninesy \scriptfont2=\sevensy \scriptscriptfont2=\fivesy
\textfont3=\tenex \scriptfont3=\tenex \scriptscriptfont3=\tenex
\textfont\itfam=\eightit  \def\it{\fam\itfam\eightit} 
\textfont\slfam=\eightsl \def\sl{\fam\slfam\eightsl} 

\setbox\strutbox=\hbox{\vrule height6pt depth2pt width0pt}%
\normalbaselines \rm}

\def\ninepoint{\def\rm{\fam0\ninerm} 
\textfont0=\ninerm \scriptfont0=\sevenrm \scriptscriptfont0=\fiverm
\textfont1=\ninei \scriptfont1=\seveni \scriptscriptfont1=\fivei
\textfont2=\ninesy \scriptfont2=\sevensy \scriptscriptfont2=\fivesy
\textfont3=\tenex \scriptfont3=\tenex \scriptscriptfont3=\tenex
\textfont\itfam=\nineit  \def\it{\fam\itfam\nineit} 
\textfont\slfam=\ninesl \def\sl{\fam\slfam\ninesl} 
\textfont\bffam=\ninebf \scriptfont\bffam=\sevenbf
\scriptscriptfont\bffam=\fivebf \def\bf{\fam\bffam\ninebf} 
\textfont\ttfam=\ninett \def\tt{\fam\ttfam\ninett} 

\normalbaselineskip=11pt
\setbox\strutbox=\hbox{\vrule height8pt depth3pt width0pt}%
\let \smc=\sevenrm \let\big=\ninebig \normalbaselines
\parindent=1em
\rm}

\def\tenpoint{\def\rm{\fam0\tenrm} 
\textfont0=\tenrm \scriptfont0=\ninerm \scriptscriptfont0=\fiverm
\textfont1=\teni \scriptfont1=\seveni \scriptscriptfont1=\fivei
\textfont2=\tensy \scriptfont2=\sevensy \scriptscriptfont2=\fivesy
\textfont3=\tenex \scriptfont3=\tenex \scriptscriptfont3=\tenex
\textfont\itfam=\nineit  \def\it{\fam\itfam\nineit} 
\textfont\slfam=\ninesl \def\sl{\fam\slfam\ninesl} 
\textfont\bffam=\ninebf \scriptfont\bffam=\sevenbf
\scriptscriptfont\bffam=\fivebf \def\bf{\fam\bffam\tenbf} 
\textfont\ttfam=\tentt \def\tt{\fam\ttfam\tentt} 

\normalbaselineskip=11pt
\setbox\strutbox=\hbox{\vrule height8pt depth3pt width0pt}%
\let \smc=\sevenrm \let\big=\ninebig \normalbaselines
\parindent=1em
\rm}

\message{fin format jgr}

\hautpagegauche={\hfill\ninerm\the\auteurcourant}
\hautpagedroite={\ninerm\the\titrecourant\hfill}
\auteurcourant={R.G.\ Novikov}
\titrecourant={Explicit formulas and global uniqueness for phaseless inverse
scattering in multidimensions}

\magnification=1200
\font\Bbb=msbm10
\def\R{\hbox{\Bbb R}}
\def\C{\hbox{\Bbb C}}
\def\N{\hbox{\Bbb N}}

\def\b{\backslash}

\def\ep{\varepsilon}

\vskip 2 mm
\centerline{\bf Explicit formulas and global uniqueness for phaseless  inverse scattering}
\centerline{\bf in multidimensions}

\vskip 2 mm
\centerline{\bf R.G.\ Novikov}
\vskip 2 mm

\noindent
{\ninerm CNRS (UMR 7641), Centre de Math\'ematiques Appliqu\'ees, Ecole
Polytechnique,}

\noindent
{\ninerm 91128 Palaiseau, France;}

\noindent
{\ninerm IEPT RAS, 117997 Moscow, Russia}

\noindent
{\ninerm e-mail: novikov@cmap.polytechnique.fr}

\vskip 2 mm
{\bf Abstract.}
We consider phaseless  inverse scattering for the Schr\"odinger equation with
compactly supported potential in dimension $d\ge 2$.
We give explicit formulas for solving this problem from appropriate data at high
energies. As a corollary, we give also a global uniqueness result for this problem with appropriate data
on a fixed energy neighborhood.

\vskip 2 mm
{\bf 1. Introduction}

We consider the Schr\"odinger equation
$$-\Delta\psi +v(x)\psi=E\psi,\ \ x\in\R^d,\ d\ge 2,\ E>0,\eqno(1.1)$$
where
$$\eqalign{
&v\in  L^{\infty}(\R^d),\ \ supp\,v\subset D,\cr
&D\ \ {\rm is\ an\ open\ bounded\ domain\ in}\ \ \R^d.\cr}\eqno(1.2)$$
For equation (1.1) we consider the classical scattering solutions
$\psi^+=\psi^+(x,k)$, $x\in\R^d$, $k\in\R^d$, $k^2=E$,
 specified by the following asymptotics as $|x|\to\infty$:
$$\eqalign{
&\psi^+(x,k)=e^{ikx}+c(d,|k|){e^{i|k||x|}\over |x|^{(d-1)/2}}
f(k,|k|{x\over |x|})+O\bigl({1\over |x|^{(d+1)/2}}\bigr),\cr
&c(d,|k|)=-\pi i(-2\pi i)^{(d-1)/2}|k|^{(d-3)/2},\cr}\eqno(1.3)$$
for some  a priori unknown $f$. In addition, the function $f=f(k,l)$,
$k,l\in\R^d$,\ $k^2=l^2=E$, arising in (1.3) is the classical scattering
amplitude for equation (1.1).

In order to find $\psi^+$ and $f$ from $v$ one can use, in particular, the
Lippmann-Schwinger integral equation
$$\eqalign{
&\psi^+(x,k)=e^{ikx}+\int\limits_DG^+(x-y,k)v(y)\psi^+(y,k)dy,\cr
&G^+(x,k)=-(2\pi)^{-d}
\int\limits_{\R^d}{e^{i\xi x}d\xi\over {\xi^2-k^2-i0}},\cr}\eqno(1.4)$$
and the formula
$$f(k,l)=(2\pi)^{-d}\int\limits_De^{-ily}v(y)\psi^+(y,k)dy,\eqno(1.5)$$
where $x,k,l\in\R^d$,  $k^2=l^2=E>0$; see e.g. [BS], [F2].

The scattering amplitude $f$ for equation (1.1) at fixed  $E$ is defined on
$${\cal M}_E=\{k\in\R^d,\ l\in\R^d:\ \ k^2=l^2=E\},\ \ E>0.\eqno(1.6)$$
In addition to $f$ on ${\cal M}_E$, we consider also
$f\big|_{\Gamma_E}$, where
$$\eqalign{
&\Gamma_E=\{k=k_E(p),\ l=l_E(p):\ p\in \bar{\cal B}_{2\sqrt{E}}\},\cr
&k_E(p)={p\over 2}+\bigl(E-{p^2\over 4}\bigr)^{1/2}\gamma(p),\
l_E(p)=-{p\over 2}+\bigl(E-{p^2\over 4}\bigr)^{1/2}\gamma(p),\cr}\eqno(1.7)$$
$${\cal B}_r=\{p\in\R^d:\ |p|< r\},\ \ \bar{\cal B}_r=\{p\in\R^d:\ |p|\le r\},\ \ r>0,\eqno(1.8)$$
where $\gamma$ is a piecewise continuous
vector-function on $\R^d$ such that
$$|\gamma(p)|=1,\ \ \gamma(p)p=0,\ \  p\in\R^d.\eqno(1.9)$$
One can see that
$$\Gamma_E\subset {\cal M}_E,\  dim\,\Gamma_E=d,\ \ dim\,{\cal M}_E=2d-2,\ E>0,\ d\ge 2.\eqno(1.10)$$
Let
$${\cal M}_{\Lambda}=\cup_{\scriptstyle E\in\Lambda}{\cal M}_E,\ \ \Gamma_{\Lambda}=\cup_{\scriptstyle E\in\Lambda}\Gamma_E,
\eqno(1.11)$$
where $\Lambda\subseteq\R_+=]0,+\infty[$.

We start with the  following inverse scattering problems for equation (1.1)
under assumptions (1.2):

\vskip 2 mm
{\bf Problem 1.1.}
Reconstruct potential $v$ on $\R^d$ from its scattering amplitude  $f$ on some appropriate  ${\cal M}^{\prime}\subseteq {\cal M}_{\R_+}$.

\vskip 2 mm
{\bf Problem 1.2.}
Reconstruct potential $v$ on $\R^d$ from its phaseless scattering data  $|f|^2$ on some appropriate  ${\cal M}^{\prime}\subseteq {\cal M}_{\R_+}$.

Note that in quantum mechanical scattering  experiments (in framework of model described by equation (1.1))
the phaseless scattering data  $|f|^2$ can  be  measured directly, whereas the complete scattering amplitude $f$
is not accessible for direct measurements. Therefore, Problem 1.2 is of particular interest from applied point of view
in the framework of quantum mechanical inverse scattering.
However, in the literature much more results are given on Problem 1.1 (see [ABR],[B], [BAR], [ChS], [EW], [E],
[ER], [F1], [F2], [G], [HH], [HN], [I], [IN], [Me], [Mo], [Ne], [N1]-[N7], [R], [S], [VW], [WY]
and references therein) than on  Problem 1.2
(see chapter X of [ChS] and recent works [K1], [K2] and references therein, where in [K1], [K2] some similar problem is considered).

In particular, for Problem 1.1 it is well known that the scattering amplitude $f$ at high energies uniquely determines $v$
via the formulas
$$\eqalignno{
&\hat v(k-l)=f(k,l)+O(E^{-1/2})\ \ {\rm as}\ \  E\to +\infty,\ \ (k,l)\in {\cal M}_E,&(1.12)\cr
&\hat v(p)=(2\pi)^{-d}\int\limits_De^{ipx}v(x)dx,\ \ p\in\R^d;&(1.13)\cr}$$
see, for example, [F1], [N7].

On the other hand, for Problem 1.2 it is well known that the phaseless scattering data $|f|^2$ on ${\cal M}_{\R_+}$
do not determine $v$ uniquely, in general. In particular, we have that
$$\eqalign{
&f_y(k,l)=e^{i(k-l)y}f(k,l),\cr
&|f_y(k,l)|^2=|f(k,l)|^2,\ \ (k,l)\in {\cal M}_{\R_+},\ \ y\in\R^d,\cr}\eqno(1.14)$$
where $f$ is  the scattering amplitude for $v$ and $f_y$ is  the scattering amplitude for $v_y$,
where
$$v_y(x)=v(x-y),\ \ x\in\R^d,\ \ y\in\R^d;\eqno(1.15)$$
see, for example, Lemma 1 of [N6].

In the present work, in view of the aforementioned  nonuniqueness for Problem 1.2 we modify this problem
into Problem 1.3 formulated below. Let
$$S=\{|f|^2,|f_j|^2,\ \ j=1,\ldots,n\},\eqno(1.16)$$
where $f$ is the initial scattering amplitude for $v$ satisfying (1.2) and $f_j$ is the  scattering amplitude for
$$v_j=v+w_j,\ \  j=1,\ldots,n,\eqno(1.17)$$
where $w_1,\ldots,w_n$ are additional a priori known background scatterers such that
$$\eqalign{
&w_j\in L^{\infty}(\R^d),\ \ supp\,w_j\subset \Omega_j,\cr
&\Omega_j\ \ {\rm is\ an\ open\ bounded\ domain\ in}\ \ \R^d,\ \Omega_j\cap D=\emptyset,\cr}\eqno(1.18a)$$
$$\eqalign{
&w_j\ne 0,\ w_{j_1}\ne w_{j_2}\ \ {\rm for}\ \ j_1\ne j_2\ \ {\rm in}\ \ L^{\infty}(\R^d),\cr
&j,\ j_1,\ j_2\in \{1,\ldots,n\}.\cr}\eqno(1.18b)$$

\vskip 2 mm
{\bf Problem 1.3.}
Reconstruct potential $v$ on $\R^d$ from the phaseless scattering data  $S$ on some appropriate  ${\cal M}^{\prime}\subseteq {\cal M}_{\R_+}$
and for some appropriate background scatterers $w_1,\ldots,w_n$.

Note also that Problems 1.1, 1.2, 1.3 can be considered as examples of ill-posed problems;
see [LRS] for an introduction to this theory.

Problem 1.3 in dimension $d=1$ was, actually, considered in [AS] for $n=1$. However, to our knowledge,
Problem 1.3 in dimension $d\ge 2$ was not yet considered in the literature before the present work.

Results of the present work can be summarized as follows.

First, we give explicit formulas for solving Problem 1.3 in dimension  $d\ge 2$ for $n=2$ and
${\cal M}^{\prime}=\Gamma_{\Lambda}$ defined by (1.7), (1.11) for any unbounded $\Lambda\subset \R_+$;
see Theorem 2.1, Remark 3.1 and Corollary 2.1 of Section 2. As an example of $\Lambda$ for this result one can take
$\Lambda=[E_0,+\infty[,E_0>0$, or just $\Lambda$ of Remark 2.1.

Second,  we give a global uniqueness result for Problem 1.3 in dimension  $d\ge 2$ for $n=2$ and
${\cal M}^{\prime}=\Gamma_{\Lambda}$ for any bounded infinite $\Lambda\subset \R_+$;
see Theorem 2.2 of Section 2. As an example of $\Lambda$ for this result one can take
$\Lambda=]E_0-\ep,E_0+\ep [,E_0>0,\ \ep>0,\ E_0-\ep\ge 0$, or just $\Lambda$ of Theorem 2.2.

In addition, we indicate possible extensions of the aforementioned results to the case $n=1$;
see Propositions 2.1, 2.2 of Section 2.

The progress of the present work in comparison with the recent works [K1], [K2] includes explicit
formulas for phaseless inverse scattering at high energies and no  assumption that $v\ge 0$.
 In addition, in the present work we consider inverse scattering from far field phaseless  scattering data
 (and not from near field phaseless  scattering data as in [K1], [K2]).

The main statements of the present work are presented in detail in the next section.

\vskip 2 mm
{\bf 2. Main statements}
\vskip 2 mm
{\it 2.1. Notations and related remarks.}
Let
$$\hat u(p)=(2\pi)^{-d}\int\limits_{\R^d}e^{ipx}u(x)dx,\ \ p\in\R^d,\eqno(2.1)$$
where $u$ is a test function on $\R^d$. In particular, we consider $\hat u=\hat v$, $\hat w_j$ for
$u=v$, $w_j$, $j=1,\ldots,n$, where $v$, $w_j$ satisfy (1.2), (1.18).

Note that if
$$u_y(x)=u(x-y),\ \ x,y\in\R^d,\eqno(2.2a)$$
then
$$\hat u_y(p)=e^{ipy}\hat u(p),\ \ p\in\R^d.\eqno(2.2b)$$
We represent $\hat v$ and $\hat w_j$ as follows:
$$\eqalign{
&\hat v(p)=|\hat v(p)|\theta(p),\ \ \theta(p)=e^{i\alpha(p)},\cr
&\hat w_j(p)=|\hat w_j(p)|\omega_j(p),\ \ \omega_j(p)=e^{i\beta_j(p)},\cr}\eqno(2.3)$$
where $p\in\R^d$, $j=1,\ldots,n$.

We consider the following sets:
$$\eqalignno{
&A_y=\{p\in\R^d:\ \ e^{2ipy}=1\},\ \ y\in\R^d,&(2.4)\cr
&Z_0=\{p\in\R^d:\ \ |\hat v(p)|=0\},\ Z_j=\{p\in\R^d:\ \ |\hat w_j(p)|=0\},\ \ j=1,\ldots,n,&(2.5)\cr
&Y_{j_1,j_2}=\{p\in\R^d\b (Z_{j_1}\cup Z_{j_2}):(\omega_{j_1}(p))^2=(\omega_{j_2}(p))^2\},\ \ 1\le j_1,j_2\le n,\ j_1\ne j_2.&(2.6)\cr}$$

We have, in particular, that
$$A_y\ \ {\rm is\ closed\ and}\ \ Mes\,A_y=0\ \ {\rm in}\ \ \R^d,\ y\ne 0.\eqno(2.7)$$

Assumptions (1.2) on $v$ imply, in particular, that $\hat v$ is (complex-valued) real-analytic
on $\R^d$. Therefore:
$$Z_0\ \ {\rm is\ closed\ in}\ \ \R^d;\ \ Mes\,Z_0=0\ \ {\rm in}\ \ \R^d\ \ {\rm if}\ \ \hat v\not\equiv 0.\eqno(2.8)$$

Assumptions (1.18) on $w_j$ imply, in particular, that $\hat w_j$ is (complex-valued) real-analytic
on $\R^d$ and $\hat w_j\not\equiv 0$,\ \ $j=1,\ldots,n$. Therefore,
$$Z_j\ \ {\rm is\ closed\ and}\ \ Mes\,Z_j=0\ \ {\rm in}\ \ \R^d,\ j=1,\ldots,n.\eqno(2.9)$$
In addition, if
$$w_j(x)=w_j^0(|x-y|),\ \ x\in\R^d,\ \ {\rm for\ some}\ \ w_j^0,\eqno(2.10)$$
for some $j$ and some $y\in\R^d$,  then
$$Z_j=\{p\in\R^d:\ \ |p|\in {\cal R}_j\},\eqno(2.11)$$
where ${\cal R}_j$ is a discrete set in $\R_+$ without accumulation points (except $+\infty$) and
${\cal R}_j$ is independent of $y$.

In addition, taking into account (2.2), if
$$w_{j_2}(x)=w_{j_1}(x-y),\ \ x\in\R^d,\ \ j_2\ne j_1,\eqno(2.12)$$
for some $j_1$, $j_2$ and some $y\in\R^d\b \{0\}$, then
$$Y_{j_1,j_2}\subseteq A_y.\eqno(2.13)$$

\vskip 2 mm
{\it 2.2. Results on Problem 1.3 in dimension $d\ge 2$ for $n=2$.}

{\bf Theorem 2.1.}
{\sl
Suppose that complex-valued $v$ satisfies (1.2), complex-valued $w_j$ satisfies (1.18a), $j=1,2$, $d\ge 2$.
Then the following formulas hold:
$$|\hat v_j(p)|^2=\lim_{\scriptstyle p=k-l,(k,l)\in {\cal M}_E, \atop\scriptstyle E\to +\infty}
|f_j(k,l)|^2\ \ {\rm for\ each}\ \ p\in\R^d,\ j=0,1,2,\eqno(2.14)$$
$$\eqalign{
&||\hat v_j(p)|^2-|f_j(k,l)|^2|\le c(D_j)N_j^3E^{-1/2},\cr
&p=k-l,\ \ (k,l)\in {\cal M}_E,\ E^{1/2}\ge\rho(D_j,N_j),\ j=0,1,2,\cr}\eqno(2.15)$$
where $v_0=v$, $f_0=f$, $D_0=D$, $v_j$ is defined by (1.17) and $D_j=D\cup\Omega_j$ for $j=1,2$,
$\|v_j\|_{L^{\infty}(D_j)}\le N_j$, $j=0,1,2$, and $c$, $\rho$ are defined by (3.10), (3.11).
Suppose, in addition, that $w_1$, $w_2$ satisfy (1.18b) and that
$$Mes\,\bar Y_{1,2}=0\ \ {\rm in}\ \ \R^d,\eqno(2.16)$$
where $\bar Y_{1,2}$ denotes the closure of $Y_{1,2}$ in $\R^d$. Then the following formula holds:
$$\eqalign{
&\pmatrix{
\cos\alpha\cr
\sin\alpha\cr}=(\sin(\beta_2-\beta_1))^{-1}\times\cr
&\pmatrix{
\sin\beta_2\ &\ -\sin\beta_1\cr
-\cos\beta_2\ &\ \cos\beta_1\cr}
\pmatrix{
(2|\hat v||\hat w_1|)^{-1}(|\hat v_1|^2-|\hat v|^2-|\hat w_1|^2)\cr
(2|\hat v||\hat w_2|)^{-1}(|\hat v_2|^2-|\hat v|^2-|\hat w_2|^2)\cr},\cr}\eqno(2.17)$$
$\alpha=\alpha(p)$, $|\hat v|=|\hat v(p)|$, $\beta_j=\beta_j(p)$, $|\hat w_j|=|\hat w_j(p)|$, $j=1,2$,
$p\in\R^d\b (Z_0\cup Z_1\cup Z_2\cup\bar Y_{1,2})$, where $\alpha$, $\beta_1$, $\beta_2$ are defined in (2.3).

}

Theorem 2.1 is proved in Section 3.

\vskip 2 mm
{\bf Remark 2.1.}
Formulas (2.14), (2.15) of Theorem 2.1 remain valid with $\Gamma_E$ in place of ${\cal M}_E$,
where $\Gamma_E$ is defined by (1.7). In addition, taking into account (2.9), (2.16) these formulas
can be considered as explicit formulas for finding $\hat v$ on $\R^d$ from $S=\{|f|^2,|f_1|^2,|f_2|^2\}$
on $\Gamma_{\Lambda}$ and background $w_1$, $w_2$ for any
$$\Lambda=\{E_j\in\R_+:\ \ j\in\N,\ \ E_j\to\infty\ \ {\rm as}\ \ j\to\infty\},\eqno(2.18)$$
where $\Gamma_{\Lambda}$ is defined in (1.11).

\vskip 2 mm
{\bf Corollary 2.1.}
{\sl
Let all assumptions of Theorem 2.1 on $v$ and $w_1$, $w_2$ be fulfilled. Let $\Lambda$
be defined as in (2.18). Then $S=\{|f|^2,|f_1|^2,|f_2|^2\}$ on $\Gamma_{\Lambda}$ and
background $w_1$, $w_2$ uniquely determine $v$ in $L^{\infty}(\R^d)$ via formulas (2.14), (2.15)
 and the inverse Fourier transform.
 }

In addition to results of Theorem 2.1, Remark 2.1 and Corollary 2.1 on the explicit reconstruction
from phaseless scattering data at high energies, we have also the following global uniqueness result
for the case of finite energies:

{\bf Theorem 2.2.}
{\sl
Let  $v$ satisfy (1.2),  $w_1$, $w_2$ satisfy (1.18), (2.16), $d\ge 2$, and $v$, $w_1$, $w_2$
be real-valued. Let
$$\Lambda=\{E_j\in\R_+:\ \ j\in\N,\ \ E_{j_1}\ne E_{j_2}\ \ {\rm for}\ \ j_1\ne j_2,\ E_j\to E_*\ \ {\rm as}\ \
j\to\infty\},\ \ E_*>0.\eqno(2.19)$$
Then $S=\{|f|^2,|f_1|^2,|f_2|^2\}$ on $\Gamma_{\Lambda}$ and
background $w_1$, $w_2$ uniquely determine $v$ in $L^{\infty}(\R^d)$.
}

Theorem 2.2 is proved in Section 4.

\vskip 2 mm
{\it 2.3. Results for the case $n=1$.}
\vskip 2 mm
{\bf Proposition 2.1.}
{\sl
If complex-valued $v$ satisfies (1.2), complex-valued $w_1$ satisfies (1.18a),  $d\ge 2$, then
formulas (2.14), (2.15) hold for $j=0,1$. If, in addition, $w_1\ne 0$ in $L^{\infty}(\R^d)$,
then
$$\eqalign{
&\cos(\alpha-\beta_1)=(2|\hat v||\hat w_1|)^{-1}(|\hat v_1|^2-|\hat v|^2-|\hat w_1|^2),\cr
&\alpha=\alpha(p),\ |\hat v|=|\hat v(p)|,\ \beta_1=\beta_1(p),\ |\hat w_1|=|\hat w_1(p)|,\
p\in\R^d\b (Z_0\cup Z_1),\cr}\eqno(2.20)$$
where $\alpha$, $\beta_1$ are defined in (2.3).
}

Proposition 2.1 is proved in Section 3.

\vskip 2 mm
{\bf Proposition 2.2.}
{\sl
(A) There are not more than two different complex-valued potentials $v$  satisfying (1.2) with given
$S=\{|f|^2,|f_1|^2\}$ on $\Gamma_{\Lambda}$ and background complex-valued $w_1$ satisfying (1.18a),
$w_1\ne 0$ in $L^{\infty}(\R^d)$, where $\Lambda$ is defined as in (2.18).
(B) There are not more than two different real-valued potentials $v$  satisfying (1.2) with given
$S=\{|f|^2,|f_1|^2\}$ on $\Gamma_{\Lambda}$ and background real-valued $w_1$ satisfying (1.18a),
$w_1\ne 0$ in $L^{\infty}(\R^d)$, where $\Lambda$ is defined as in (2.19).
}

Proposition 2.2 is proved in Section 5.

\vskip 2 mm
{\bf 3. Proofs of Proposition 2.1 and Theorem 2.1}
\vskip 2 mm
{\it 3.1. Preliminaries.}
Let
$$\eqalign{
&L^{\infty}_{\sigma}(\R^d)=\{u\in L^{\infty}(\R^d):\ \ \|u\|_{\sigma}< +\infty\},\cr
&\|u\|_{\sigma}=ess\,\sup\limits_{x\in\R^d}(1+|x|^2)^{\sigma/2}|u(x)|,\ \ \sigma\ge 0.\cr}\eqno(3.1)$$

Note that
$$v,\ w_j,\ v_j\in L^{\infty}_{\sigma}(\R^d)\ \ {\rm for\ each}\ \ \sigma\ge 0,\eqno(3.2)$$
where $v$,\ $w_j$,\ $v_j$, $j\in \{1,2\}$, are the potentials of Proposition 2.1 and Theorem 2.1.

We recall that
$$\eqalign{
&\|<x>^{-s}G^+(k)<x>^{-s}\|_{L^2(\R^d)\to L^2(\R^d)}\le a_0(d,s)|k|^{-1},\cr
&k\in\R^d,\ \ |k|\ge 1,\ \ {\rm for}\ \ s>1/2,\cr}\eqno(3.3)$$
where $G^+(k)$ denotes the integral operator with the Schwartz kernel $G^+(x-y,k)$ of (1.4),
$<x>$ denotes the multiplication operator by the function $(1+|x|^2)^{1/2}$;
see [E], [J] and references therein.

We will use the following detailed version of formula (1.12):
$$\eqalign{
&|f(k,l)-\hat v(k-l)|\le 2(2\pi)^{-d}a_0(d,\sigma/2)(c_1(d,\sigma)\|v\|_{\sigma})^2E^{-1/2},\cr
&(k,l)\in {\cal M}_E,\ \ E^{1/2}\ge\rho_1(d,\sigma,\|v\|_{\sigma}),\ \sigma>d,\cr}\eqno(3.4)$$
where $a_0(d,s)$ is the constant of (3.3),
$$\eqalignno{
&c_1(d,\sigma)=\bigl(\int\limits_{\R^d}{dx\over (1+|x|^2)^{\sigma/2}}\bigr)^{1/2},&(3.5)\cr
&\rho_1(d,\sigma,R)=\max(2a_0(d,\sigma/2)R,1);&(3.6)\cr}$$
see formula (2.11) of [N7].

\vskip 2 mm
{\it 3.2. Proof of formulas (2.14), (2.15).}
We have that
$$|\hat v(k-l)|\buildrel (3.1),(3.5) \over \le (2\pi)^{-d}\|v\|_{\sigma}(c_1(d,\sigma))^2,\eqno(3.7)$$
$$\eqalign{
&||f(k,l)|^2-|\hat v(k-l)|^2|=||f(k,l)|-|\hat v(k-l)||(|f(k,l)|+|\hat v(k-l)|)\le\cr
&|f(k,l)-\hat v(k-l)|(2|\hat v(k-l)|+|f(k,l)-\hat v(k-l)|),\cr}\eqno(3.8)$$
where $(k,l)\in {\cal M}_E$, $\sigma>d$. Due to  (3.4), (3.7), (3.8), we have that
$$\eqalign{
&||f(k,l)|^2-|\hat v(k-l)|^2|\le 3(2\pi)^{-d}\|v\|_{\sigma}(c_1(d,\sigma))^2|f(k,l)-\hat v(k-l)|\le\cr
&6(2\pi)^{-2d}a_0(d,\sigma/2)((c_1(d,\sigma))^4(\|v\|_{\sigma})^3E^{-1/2},\cr}\eqno(3.9)$$
$(k,l)\in {\cal M}_E$, $E^{-1/2}\ge \rho_1(d,\sigma,\|v\|_{\sigma})$, $\sigma>d$.
Formulas (2.14), (2.15) follow from (3.9) for $v=v_j$, $f=f_j$ and from the possibility of choice of
$k=k_E(p)$, $l=l_E(p)$ as in (1.7) for $d\ge 2$. In addition,
$$\eqalignno{
&c(D)=6(2\pi)^{-2d}a_0(d,\sigma/2)(c_1(d,\sigma))^4(c_2(d,\sigma))^3,&(3.10)\cr
&\rho(D,N)=\rho_1(d,\sigma,c_2(D,\sigma)N),&(3.11)\cr}$$for some fixed $\sigma>d$, where
$$c_2(D,\sigma)=\sup\limits_{x\in D}(1+|x|^2)^{\sigma/2}.\eqno(3.12)$$

\vskip 2 mm
{\it 3.3. Proof of formula (2.20).}
We have that
$$\eqalign{
&|\hat v_1|^2\buildrel (1.17) \over =|\hat v+\hat w_1|^2\buildrel (2.3) \over =||\hat v|e^{i\alpha}+|\hat w_1|e^{i\beta_1}|=\cr
&(|\hat v|\cos\alpha+|\hat w_1|\cos\beta_1)^2+(|\hat v|\sin\alpha+|\hat w_1|\sin\beta_1)^2=\cr
&|\hat v|^2+|\hat w_1|^2+2|\hat v||\hat w_1|(\cos\alpha\cos\beta_1+\sin\alpha\sin\beta_1)\ \ {\rm on}\ \ \R^d.\cr}\eqno(3.13)$$
Formula (2.20) follows from (3.13).

\vskip 2 mm
{\it 3.4. Proof of formula (2.17).}
Using (3.13) and analogous formula for $\hat v_2=\hat v+\hat w_2$, we obtain the system

$$\eqalign{
&\pmatrix{
\cos\beta_1\ &\ \sin\beta_1\cr
\cos\beta_2\ &\ \sin\beta_2\cr}
\pmatrix{
\cos\alpha\cr
\sin\alpha\cr}=
\pmatrix{
(2|\hat v||\hat w_1|)^{-1}(|\hat v_1|^2-|\hat v|^2-|\hat w_1|^2)\cr
(2|\hat v||\hat w_2|)^{-1}(|\hat v_2|^2-|\hat v|^2-|\hat w_2|^2)\cr},\cr}\eqno(3.14)$$
on $\R^d\b (Z_0\cup Z_1\cup Z_2)$.

Formula (2.17) follows from (3.14).

\vskip 2 mm
{\it 3.5. Final remark.}
Proposition 2.1 and Theorem 2.1 follow from formulas (2.14), (2.15), (2.20), (2.17) proved in
Subsections 3.2, 3.3, 3.4.

\vskip 2 mm
{\bf 4. Proof of Theorem 2.2}

Let
$$\Delta_{E_0,E}=\{(k,l)\in\Gamma_E:\ \ k-l\in {\cal B}_{2\sqrt{E_0}}\},\ \ 0<E_0\le E,\eqno(4.1)$$
where $\Gamma_E$, ${\cal B}_r$ are defined by (1.7), (1.8).

Theorem 2.2 follows from:

(1) the formulas of Theorem 2.1 with $\Delta_{E_*,E}$ in place of ${\cal M}_E$
and $|p|<2\sqrt{E_*}$,

(2) the fact that $\hat v$ on ${\cal B}_{2\sqrt{E_*}}\b (Z_1\cup Z_2\cup \bar Y_{1,2})$
uniquely determines $\hat v$ on $\R^d$ (since $\hat v$ is real-analytic on $\R^d$), and

(3) the results of Lemma 4.1 for $v$ and for $v=v+w_j$, $j=1,2$.

\vskip 2 mm
{\bf Lemma 4.1.}
{\sl
Let $v$ satisfy (1.2) and be real-valued. Then:
\item{(a)} $|f(k_E(p),l_E(p))|^2$ is real-analytic
in $E\in ]p^2/4,+\infty[$ for fixed $p\in\R^d$, where $k_E(p)$, $l_E(p)$ are defined in (1.7 );
\item{(b)} $|f|^2$ on $\Gamma_{\Lambda}$ uniquely determines $|f|^2$ on $\Delta_{E_*,E}$ for each $E\ge E_*$,
where $\Gamma_{\Lambda}$, $\Lambda$ are defined in (1.11), (2.19).
}

Statement (b) of Lemma 4.1 follows from statement  (a) of Lemma 4.1 and the property that the
accumulation point $E_*\in ]p^2/4, +\infty[$ if $p\in {\cal B}_{2\sqrt{E_*}}$.

In turn, statement  (a)  of Lemma 4.1 follows from the presentation
$$|f|^2=f\bar f \eqno(4.2)$$
and from Lemma 4.2.

\vskip 2 mm
{\bf Lemma 4.2.}
{\sl
Let $v$ satisfy (1.2) and be real-valued. Then $f(k_E(p),l_E(p))$ admits holomorphic extension in $E$
to an open ${\cal N}$ in $\C$, where $]p^2/4, +\infty[\subset {\cal N}$, at fixed $p\in\R^d$.
}

Lemma 4.2 follows from:
\item{(1)} the integral equation (1.4) for $\psi^+$ on $D$ and the presentation (1.5) for $f$, where
\item{   } $k=k_E(p)$, $l=l_E(p)$;
\item{(2)} the property that
$$G^+(x,k)= G^+_0(|x|,|k|),\ \ x\in\R^d,\ \ k\in\R^d\b \{0\},\eqno(4.3)$$
where $G^+$ is the function of (1.4) and $G^+_0$ depends also on $d$;
\item{(3)} the properties that:
$|k_E(p)|=E^{1/2}$ for $E\in ]p^2/4, +\infty[$, $p\in\R^d$;
$E^{1/2}$ is holomorphic in $E\in\C\b ]-\infty,0]$; $(E-p^2/4)^{1/2}$ is holomorphic in $E\in\C\b ]-\infty,p^2/4]$;
$p\in\R^d$; $G_0^+(r,\kappa)$ is holomorphic in $\kappa\in\C$ for odd $d\ge 3$ and in $\kappa\in\C\b ]-\infty,0]$
for even $d\ge 2$, where $r>0$;
\item{(4)} the result that (1.4) with $k=k_E(p)$ is a Fredholm integral equation of the second kind for
$\psi^+(\cdot,k)\in L^2(D)$ with holomorphic dependence on the parameter
\item{   } $E\in\C\b ]-\infty,p^2/4]$ at fixed $p\in\R^d$;
\item{(5)} the result that (1.4) is uniquely solvable for $\psi^+(\cdot,k)\in L^2(D)$ for each $k\in\R^d\b \{0\}$
under our assumptions on $v$.

In connection with basic properties of function $G^+$ and basic properties of the

\noindent
 Lippmann-Schwinger integral equation (1.4) we refer also to  [BS], [F2], [Me].

 \vskip 2 mm
 {\bf 5. Proof of Proposition 2.2}
 \vskip 2 mm
 {\it Proof of part (A).}
 Due to formulas (2.14), (2.15) with $\Gamma_E$ in place of ${\cal M}_E$, we have that
  $S=\{|f|^2,|f_1|^2\}$ on $\Gamma_{\Lambda}$ uniquely determine $|\hat v|$, $|\hat v_1|$ on $\R^d$. If $|\hat v|\equiv 0$, then $v=0$ in
 $L^{\infty}(\R^d)$. Therefore, it remains to consider the case when $|\hat v|\not\equiv 0$.

 Due to (2.8), (2.9), $j=1$, and continuity of $\hat v$, $\hat w_1$, we can choose $p^{\prime}\in\R^d$, $r^{\prime}>0$
 such that
 $$\eqalignno{
 &|\hat v(p)|\ne 0,\ |\hat w_1(p)|\ne 0\ \ {\rm for}\ \ p\in {\cal B}_{p^{\prime},r^{\prime}},&(5.1)\cr
 &{\cal B}_{p^{\prime},r^{\prime}}=\{p\in\R^d:\ \ |p-p^{\prime}|<r^{\prime}\}.&(5.2)\cr}$$
 Therefore, formula (2.20) for $\cos(\alpha-\beta_1)$ holds for each $p\in {\cal B}_{p^{\prime},r^{\prime}}$.

 If $\cos(\alpha-\beta_1)\equiv 1$ on ${\cal B}_{p^{\prime},r^{\prime}}$, then $\alpha\equiv\beta_1$ ($mod\,2\pi$)
on ${\cal B}_{p^{\prime},r^{\prime}}$.
If $\cos(\alpha-\beta_1)\equiv -1$ on ${\cal B}_{p^{\prime},r^{\prime}}$, then $\alpha\equiv\beta_1 +\pi$ ($mod\,2\pi$)
on ${\cal B}_{p^{\prime},r^{\prime}}$. And in both cases $\hat v=|\hat v|e^{i\alpha}$ is uniquely determined on
${\cal B}_{p^{\prime},r^{\prime}}$ by $|\hat v|$, $|\hat v_1|$, $\hat w_1=|\hat w_1|e^{i\beta_1}$ on ${\cal B}_{p^{\prime},r^{\prime}}$.

Due to continuity of $e^{i\alpha}$, $e^{i\beta_1}$ on ${\cal B}_{p^{\prime},r^{\prime}}$ we can choose
$p^{\prime\prime}\in {\cal B}_{p^{\prime},r^{\prime}}$ and $r^{\prime\prime}\in ]0,r^{\prime}[$ such that
$$-1<c_{min}\le\cos(\alpha-\beta_1)\le c_{max}<1\ \ {\rm on}\ \ {\cal B}_{p^{\prime\prime},r^{\prime\prime}}\eqno(5.3)$$
for some fixed $c_{min}$, $c_{max}$. Therefore, due to formula (2.20) for $\cos(\alpha-\beta_1)$ and continuity of
$\alpha$, $\beta_1$ ($mod\,2\pi$) on ${\cal B}_{p^{\prime\prime},r^{\prime\prime}}$ we have that either
$$\alpha=\beta_1+arc\cos((2|\hat v||\hat w_1|)^{-1}(|\hat v_1|^2-|\hat v|^2-|\hat w_1|^2))\eqno(5.4a)$$
or
$$\alpha=\beta_1-arc\cos((2|\hat v||\hat w_1|)^{-1}(|\hat v_1|^2-|\hat v|^2-|\hat w_1|^2))\eqno(5.4b)$$
($mod\,2\pi$) on ${\cal B}_{p^{\prime\prime},r^{\prime\prime}}$, where $arc\cos$ takes values in $[0,\pi]$.
Therefore, there are not more than two different $\hat v=|\hat v|e^{i\alpha}$ on ${\cal B}_{p^{\prime\prime},r^{\prime\prime}}$
with given  $|\hat v|$, $|\hat v_1|$, $\hat w_1=|\hat w_1|e^{i\beta_1}$ on ${\cal B}_{p^{\prime\prime},r^{\prime\prime}}$.
In turn, $\hat v$ on ${\cal B}_{p^{\prime\prime},r^{\prime\prime}}$ uniquely determines $\hat v$ on $\R^d$ due to real
analyticity of $\hat v$.

This completes the proof of part (A) of Proposition 2.2.

{\it Proof of part (B).}
Due to statement (b) of Lemma 4.1 (for $v$ and for $v=v+w_1$), $S=\{|f|^2,|f_1|^2\}$ on $\Gamma_{\Lambda}$ uniquely determine
$S$ on $\Delta_{E_*,E}$. Due to formulas (2.14), (2.15) with $\Delta_{E_*,E}$ in place of ${\cal M}_E$ and
$|p|<2\sqrt{E_*}$, we have that $S$ on $\Delta_{E_*,E}$  uniquely determine
$|\hat v|$, $|\hat v_1|$ on ${\cal B}_{2\sqrt{E_*}}$.

Then in a completely similar way with the proof of part (A) of Proposition 2.2 we obtain that there are not
more than two different $\hat v$ on ${\cal B}_{2\sqrt{E_*}}$ with given $|\hat v|$, $|\hat v_1|$, $\hat w_1$
 on ${\cal B}_{2\sqrt{E_*}}$.

 Finally, $\hat v$ on ${\cal B}_{2\sqrt{E_*}}$  uniquely determines  $\hat v$ on $\R^d$ due to real analyticity of
 $\hat v$.

 This completes the proof of part (B).

\vskip 4 mm
{\bf References}
\vskip 2 mm
\item{[ AS]} T. Aktosun, P.E. Sacks, Inverse problem on the line without phase information,
Inverse Problems 14, 1998, 211-224.
\item{[ABR]} N.V. Alexeenko, V.A. Burov, O.D. Rumyantseva, Solution of the
 three-dimensional
acoustical inverse scattering problem. The modified Novikov
algorithm, Acoust. J. 54(3), 2008, 469-482 (in Russian), English transl.:
Acoust. Phys. 54(3), 2008, 407-419.
\item{[ BS]} F.A. Berezin, M.A. Shubin, The Schr\"odinger Equation,
Vol. 66 of Mathematics and Its Applications, Kluwer Academic, Dordrecht, 1991.
\item{[  B]} A.L. Buckhgeim, Recovering a potential from Cauchy data in the
two-dimensional case, J. Inverse Ill-Posed Probl. 16(1), 2008,  19-33.
\item{[BAR]} V.A. Burov, N.V. Alekseenko, O.D. Rumyantseva, Multifrequency
generalization
of the Novikov algorithm for the two-dimensional inverse scattering
problem, Acoust. J. 55(6), 2009, 784-798 (in Russian); English transl.:
 Acoustical Physics 55(6), 2009,  843-856.
\item{[ChS]} K. Chadan, P.C. Sabatier, Inverse Problems in Quantum Scattering
Theory, 2nd edn. Springer, Berlin, 1989
\item{[ EW]} V.Enss, R.Weder, Inverse potential scattering: a geometrical approach,
Mathematical quantum theory. II. Schr\"odinger operators (Vancouver, BC, 1993), 151-162,
CRM Proc. Lecture Notes, 8, Amer.Math.Soc., Providence, RI, 1995.
\item{[  E]} G. Eskin, Lectures on Linear Partial Differential Equations,
Graduate Studies in Mathematics, Vol.123, American Mathematical Society, 2011.
\item{[ ER]} G. Eskin, J. Ralston, Inverse backscattering problem in
three dimensions, Commun. Math. Phys. 124, 1989, 169-215.
\item{[ F1]} L.D. Faddeev, Uniqueness of the solution of the inverse
scattering problem, Vest. Leningrad Univ. 7, 1956, 126-130 [in Russian].
\item{[ F2]} L.D. Faddeev, The inverse problem in the quantum theory of
scattering.II, Current problems in mathematics, Vol. 3, 1974, pp. 93-180,
259. Akad.
Nauk SSSR Vsesojuz. Inst. Naucn. i Tehn. Informacii, Moscow(in Russian);
English transl.: J.Sov. Math. 5, 1976, 334-396.
\item{[  G]} P.G. Grinevich, The scattering transform for the two-dimensional
Schr\"odinger operator with a potential that decreases at infinity at fixed
nonzero energy, Uspekhi Mat. Nauk 55:6(336),2000, 3-70 (Russian); English
translation: Russian Math. Surveys 55:6, 2000, 1015-1083.
\item{[ HH]} P. H\"ahner, T. Hohage, New stability estimates for the inverse
acoustic inhomogeneous
medium problem and applications, SIAM J. Math. Anal., 33(3),
2001, 670-685.
\item{[ HN]} G.M. Henkin, R.G. Novikov, The $\bar\partial$-equation in the
multidimensional
inverse scattering problem, Uspekhi Mat. Nauk 42(3), 1987, 93-152
(in Russian);
English transl.: Russ. Math. Surv. 42(3), 1987, 109-180.
\item{[  I]} M.I. Isaev, Exponential instability in the inverse scattering
problem on the
energy interval, Funkt. Anal. Prilogen. 47(3), 2013, 28-36 (in Russian);
English transl.: Funct. Anal. Appl. 47(3), 2013, 187-194.
\item{[IN]} M.I. Isaev, R.G. Novikov, New global stability estimates for
monochromatic
inverse acoustic scattering, SIAM J. Math. Anal. 45(3), 2013, 1495-1504
\item{[  J]} A. Jensen, High energy resolvent estimates for generalized
 many-body Schr\"odinger operators, Publ. RIMS Kyoto Univ. 25, 1989, 155-167.
\item{[ K1]} M.V. Klibanov, Phaseless inverse scattering problems in three dimensions,
SIAM J. Appl. Math. 74, 2014, 392-410.
\item{[ K2]} M.V. Klibanov, On the first solution of a long standing problem:
uniqueness of the phaseless quantum inverse scattering problem in 3-d,
Appl. Math. Lett. 37, 2014, 82-85.
\item{[LRS]} M.M Lavrentev, V.G.Romanov, S.P. Shishatskii, Ill-posed problems of mathematical
physics and analysis, Translated from the Russian by J.R.Schulenberger. Translation edited by
Levi J.Leifman. Translations of Mathematical Monographs, 64. American Mathematical Society,
Providence, RI, 1986.
\item{[ Me]} R.B. Melrose, Geometric scattering theory, Cambridge University Press, 1995.
\item{[ Mo]} H.E. Moses, Calculation of the scattering potential from reflection coefficients,
Phys. Rev. 102, 1956, 559-567.
\item{[ Ne]} R.G.Newton, Inverse Schr\"odinger scattering in three dimensions, Springer, Berlin, 1989.
\item{[ N1]} R.G. Novikov, Multidimensional inverse spectral problem for the
equation
$-\Delta\psi+(v(x)-Eu(x))\psi=0$, Funkt. Anal. Prilozhen. 22(4), 1988, 11-22
(in Russian); English transl.: Funct. Anal. Appl. 22, 1988, 263-272.
\item{[ N2]} R.G. Novikov, The inverse scattering problem at fixed energy
level for the two-dimensional Schr\"odinger operator, J. Funct. Anal., 103,
1992, 409-463.
\item{[ N3]} R.G. Novikov, The inverse scattering problem at fixed energy for
Schr\"odinger equation with an exponentially decreasing potential,
Comm. Math.
\item{     } Phys., 161, 1994,  569-595.
\item{[ N4]} R.G. Novikov, On determination of the Fourier transform of a potential
from the scattering amplitude, Inverse Problems 17, 2001, 1243-1251.
\item{[ N5]} R.G. Novikov, The $\bar\partial$-approach to monochromatic inverse
 scattering in three dimensions, J. Geom. Anal. 18, 2008, 612-631.
\item{[ N6]} R.G. Novikov, Absence of exponentially localized solitons for the Novikov-Veselov
equation at positive energy, Physics Letters A 375, 2011, 1233-1235.
\item{[ N7]} R.G. Novikov, An iterative approach to non-overdetermined inverse scattering at
fixed energy, Mat. Sb. (to appear)
\item{[  R]} T. Regge, Introduction to complex orbital moments, Nuovo Cimento 14, 1959,
951-976
\item{[  S]} P. Stefanov, Stability of the inverse problem in potential
scattering at fixed
energy, Annales de l'Institut Fourier, tome 40(4), 1990, 867-884.
\item{[ VW]} A.Vasy, X.-P. Wang, Inverse scattering with fixed energy for
dilation-analytic potentials, Inverse Problems, 20, 2004, 1349-1354.
\item{[ WY]} R. Weder, D. Yafaev, On inverse scattering at a fixed energy
for potentials with a regular behaviour at infinity,
Inverse Problems, 21, 2005, 1937-1952.

\end